\documentclass{article}
\usepackage{spconf,amsmath,graphicx}
\usepackage{multicol}
\usepackage{multirow}
\usepackage{diagbox}
\usepackage{url}

\title{Towards Low-resource StarGAN Voice Conversion using Weight Adaptive Instance Normalization}
\name{Mingjie Chen, Yanpei Shi, Thomas Hain}
\address{Department of Computer Science, University of Sheffield, \\{mchen33, yshi30, t.hain}@sheffield.ac.uk}
\begin{document}
\ninept
\maketitle
\begin{abstract}
Many-to-many voice conversion with non-parallel training data has seen significant progress in recent years. It is challenging because of lacking of ground truth parallel data. StarGAN-based models have gained attentions because of their efficiency and effectiveness. However, most of the StarGAN-based works only focused on small number of speakers and large amount of training data. In this work, we aim at improving the data efficiency of the model and achieving a many-to-many non-parallel StarGAN-based voice conversion for a relatively large number of speakers with limited training samples. In order to improve data efficiency, the proposed model uses a speaker encoder for extracting speaker embeddings and weight adaptive instance normalization (W-AdaIN) layers. Experiments are conducted with 109 speakers under two low-resource situations, where the number of training samples is 20 and 5 per speaker. An objective evaluation shows the proposed model outperforms baseline methods significantly. Furthermore, a subjective evaluation shows that, for both naturalness and similarity, the proposed model outperforms baseline method.
\end{abstract}
\begin{keywords}
Voice Conversion, Generative Adversarial Networks, Low-resource
\end{keywords}
\section{Introduction}
\label{sec:1}
Given a voice sample of a source speaker and a voice sample of a target speaker, voice conversion aims at converting speaker properties of source sample to target speaker. Statistical models such as Gaussian mixture models (GMMs) \cite{kain1998spectral, stylianou1998continuous} have been used for voice conversion. Besides, deep neural networks (DNN) \cite{desai2010spectral,azarov2013real} have also been popular for voice conversion. However, both the GMM-based models and the DNN-based models required aligned parallel data for training, where source sample and target sample contain the same speech content information. Obtaining aligned parallel data is not easy and requires time-consuming human works.
More recently, generative models such as variational auto-encoder \cite{lopez2018information,hsu2016voice,chou2018multi} (VAE) and generative adversarial network (GAN) \cite{goodfellow2014generative} have gained attentions for non-parallel voice conversion. 

In terms of GAN-based models for non-parallel voice conversion, CycleGAN-VC \cite{kaneko2017parallel} used CycleGAN \cite{zhu2017unpaired} model. A cycle-consistency loss was used in CycleGAN-VC to avoid using aligned parallel data. StarGAN-VC \cite{kameoka2018stargan} proposed to use StarGAN \cite{choi2018stargan} model for voice conversion. It used a domain classifier module, in order to enhance the similarity of converted samples. StarGAN-VC suffered from a partial conversion issue, which means the converted voices were neutral. Also the domain classifier module influenced the voice quality. StarGAN-VC2 \cite{kaneko2019stargan} and \cite{lee2020many} were proposed to improve the performance of StarGAN-based voice conversion by removing the domain classifier module. StarGAN-VC2 proposed to use conditional instance normalization \cite{dumoulin2016learned} to improve the speaker adaptation ability of the model. However, feature-based normalization layers \cite{ulyanov2016instance,dumoulin2016learned,huang2017arbitrary} have been found causing information loss\cite{karras2020analyzing}, which could lead to low data efficiency.

Most of the mentioned StarGAN-based voice conversion works used a small number of speakers. For example, in StarGAN-VC and StarGAN-VC2, only 4 speakers were used, and the amount of the training data per speaker was 5 minutes. \cite{wang2020one} trained the StarGAN-VC model with 37 speakers, however the training data per speaker was 30 minutes in average. It is unclear whether the StarGAN-based models can keep the performance when increasing the number of speakers and decreasing the training samples.

This work aims at improving the data efficiency of the StarGAN-based model and exploring voice conversion under low-resource situations. 
We propose a weight adaptive instance normalization StarGAN-VC (WAStarGAN-VC) model. Two approaches are used to improve the data efficiency of the model: (1) unlike StarGAN-VC and StarGAN-VC2 only using speaker identity for target speaker information, we uses a speaker encoder to extract speaker embeddings from target speech; (2) instead of normalizing feature, we follow the idea from StyleGAN2 \cite{karras2020analyzing} and conduct adaptive instance normalization on the convolutional weights, to avoid information loss caused by normalization layers. The voice conversion experiments are conducted with 109 speakers under two low-resource situations. We use speaker identification and verification for objective evaluation. For subjective evaluation, we evaluate the proposed model using ABX test (similarity) and AB test (naturalness). The evaluation results show that WAStarGAN-VC outperforms the baseline models (StarGAN-VC and StarGAN-VC2). 
\section{StarGAN-based Voice Conversion}\label{sec:2}
This section reviews two previous StarGAN-based voice conversion models: StarGAN-VC \cite{kameoka2018stargan} model and StarGAN-VC2  \cite{kaneko2019stargan} model.
\subsection{StarGAN-VC Model}\label{subsec:2.1}
StarGAN-VC \cite{kameoka2018stargan} adapted and used the StarGAN \cite{choi2018stargan} model for voice conversion. 
The model is composed of three modules: a generator $G()$, a discriminator $D()$ and a domain classifier $C()$. Given a real data $x \sim p(x)$
and a target speaker identity $s_y$, the generator converts data $x$ to data $y$.
\begin{equation}\label{sec2eq:stgan_vc1_gen}
  y = G(x, s_y)  
\end{equation}
As shown in Equation \ref{sec2eq:stgan_vc1_dis}, the discriminator takes in a data $x^*$ and a speaker identity $s^*$, where $(x^*,s^*)$ can be real source data and source speaker identity $(x,s_x)$ or converted data and target speaker identity $(y,s_y)$.
\begin{equation}\label{sec2eq:stgan_vc1_dis}
    o = D(x^*, s^*),
\end{equation}
where $o$ is the output of the discriminator, $s_x$ is the source speaker id. $o$ is the probability that the input $x^*$ belongs to real data distribution. 

The loss function of StarGAN-VC has four parts:
\begin{equation}\label{sec2eq:stgan_vc1_total_loss}
    \mathcal{L}_{StarGAN-VC}^G = \mathcal{L}_{adv}^G + \mathcal{L}_{cyc}^G + \mathcal{L}_{id}^G + \mathcal{L}_{domain}^G
\end{equation}
\begin{equation}
    \mathcal{L}_{StarGAN-VC}^D = \mathcal{L}_{adv}^D
\end{equation}
The adversarial losses are defined as:
\begin{equation}\label{sec2eq:stgan_vc1_adv_loss_g}
    \mathcal{L}_{adv}^G = - \mathrm{E}_{x,s_y}[D(G(x,s_y), s_y))]
\end{equation}
\begin{equation} \label{sec2eq:stgan_vc1_adv_loss_d}
    \mathcal{L}_{adv}^D = - \mathrm{E}_{x,s_x}[D(x,s_x)] - \mathrm{E}_{x, s_y}[1 - D(G(x, s_y), s_y)]
\end{equation}
Besides, StarGAN-VC also used the identity loss $\mathcal{L}_{id}$ and the cycle consistency loss $\mathcal{L}_{cyc}$.
\begin{equation}\label{sec2eq:stgan_vc1_id_loss}
    \mathcal{L}_{id}^G= \mathrm{E}_{x, s_x}[|| x - G(x, s_x)||_1]
\end{equation}
\begin{equation}\label{sec2eq:stgan_vc1_cyc_loss}
    \mathcal{L}_{cyc}^G = \mathrm{E}_{x, s_y, s_x}[|| x - G(G(x, s_y),s_x)||_1]
\end{equation}
The domain classifier is used to force the generated data $y$ to be similar to the target speaker $s_y$.
\begin{equation}\label{sec2eq:stgan_vc1_cls_loss_c}
\mathcal{L}_{domain}^C = - \mathrm{E}_{x, s_x}[p_C(s_x|x)]
\end{equation}
\begin{equation}\label{sec2eq:stgan_vc1_cls_loss_g}
    \mathcal{L}_{domain}^G = - \mathrm{E}_{x, s_y}[p_C(s_y| G(x, s_y))]
\end{equation}
\subsection{StarGAN-VC2 Model}\label{subsec:2.2}
One of limitations of the StarGAN-VC model is that the domain classifier loss hurts the voice quality \cite{kaneko2019stargan}. Additionally, only using the target speaker identity $s_y$ in the generator and the discriminator causes the partial conversion issue \cite{kaneko2019stargan}. In order to solve the voice quality issue, the StarGAN-VC2 model removed the domain classifier module.  
Besides, to improve similarity, the StarGAN-VC2 model used the concatenation of the source speaker embedding $e_x$  and the target speaker embedding $e_y$ as speaker condition.
\begin{equation}\label{sec2:stgan_vc2_spk_id_concat}
    e_{xy} = concat([e_x, e_y])
\end{equation}
where $concat$ is the concatenation function, speaker embeddings $e_x$ an $e_y$ can be obtained through speaker ids $s_x$ and $s_y$.

StarGAN-VC2 incorporated conditional instance normalization \cite{dumoulin2016learned} (CIN) in the generator. In the StarGAN-VC2 model, CIN normalizes the feature $f$ across time and conducts affine transformation given the speaker condition $e_{xy}$.
\begin{equation}\label{sec2:stgan_vc2_cin_layer}
    CIN(f) = \gamma(e_{xy}) * (\frac{f - \mu}{\sigma}) + \beta(e_{xy}),
\end{equation}
where $CIN(f)$ is the output of CIN, $\gamma()$ and $\beta()$ are linear functions, $\mu$ and $\sigma$ are the mean and the standard deviation of the feature $f$ across time.

The training objectives of StarGAN-VC2 are similar to StarGAN-VC, including the adversarial loss, the identity loss and the cycle consistency loss. StarGAN-VC2 did not use the domain classifier loss.
\section{StarGAN Voice Conversion with Weight Adaptive Instance Normalization}
\begin{figure}[t]
    \centering
    \includegraphics[width=3in]{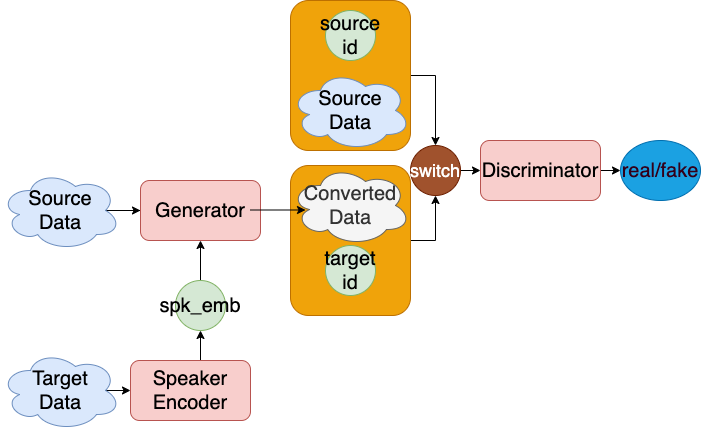}
    \label{fig:stgan3}
    \caption{Model architecture of the proposed WAStarGAN-VC model, $spk\_emb$ denotes speaker embedding}
\end{figure}
\label{sec:3}
\begin{figure}[t]
    \centering
    \includegraphics[height=3in]{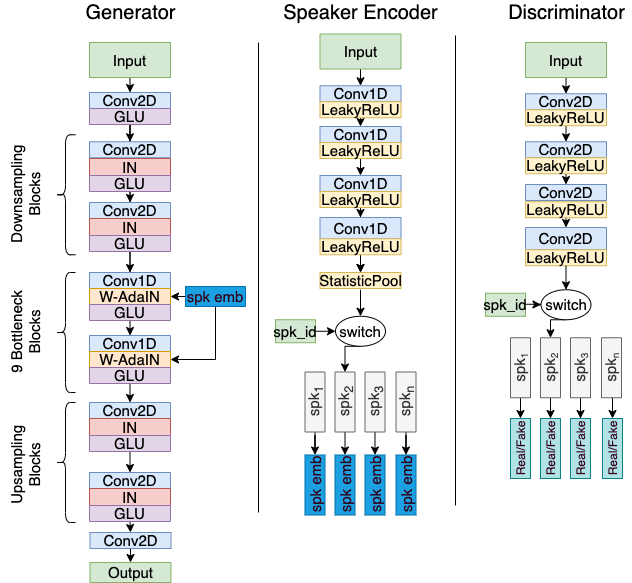}
    \caption{Module details of the proposed WAStarGAN-VC: spk\_id denotes speaker identity, spk\_emb denotes speaker embedding}
    \label{fig:stadain3}
\end{figure}
\begin{figure}[t]
    \centering
    \includegraphics[width=2in]{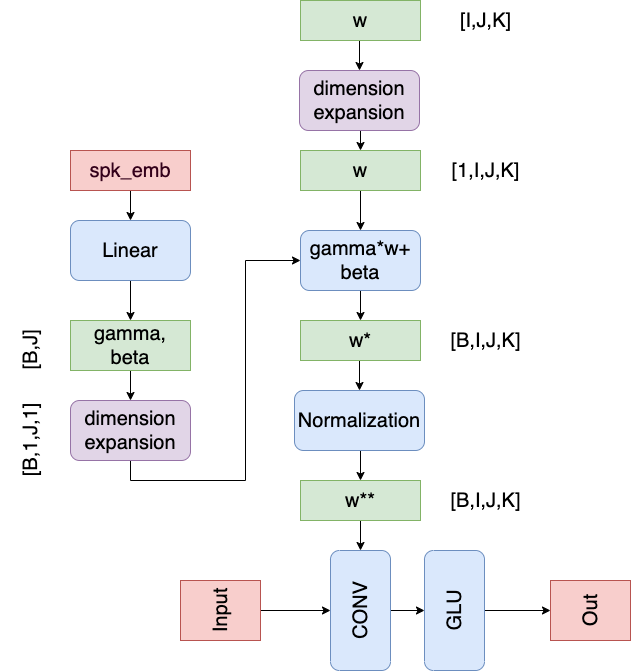}
    \caption{Weight adaptive instance normalization: $spk\_emb$ denotes speaker embedding, gamma and beta affine parameters. 'GLU' denotes activation function. $I$, $J$, $K$ are incoming channels, outcoming channels, and kernel size. $w$ is convolution kernel initialized randomly. }
    \label{fig:w_adain}
\end{figure}
Given a source data $x_s \sim p(x)$ and a target data $x_t \sim p(x)$, the proposed WAStarGAN-VC model is expected to generate a data $y_t$ that contains speech content information of $x_s$ and speaker properties of $x_t$. As shown in Figure \ref{fig:stgan3} WAStarGAN-VC is composed of three modules: a generator $G()$, a discriminator $D()$ and a speaker encoder $E()$. 

Both StarGAN-VC and StarGAN-VC2 used speaker identity as the target speaker information input. In contrast, in order to improve the data efficiency of the model, WAStarGAN-VC uses a speaker encoder to extract speaker embeddings from target data. By doing this, the model is expected to learn speaker embeddings more efficiently. On the other hand, it has been found that normalization layers such as instance normalization \cite{ulyanov2016instance} could cause information loss \cite{karras2020analyzing}. WAStarGAN-VC proposes to normalize and transform convolutional weights, to improve the data efficiency of the model as in StyleGAN2\cite{karras2020analyzing}.
\subsection{Generator with Weight Adaptive Instance Normalization}
WAStarGAN-VC uses a 2-1-2 model architecture for the generator, which is similar to CycleGAN-VC \cite{kaneko2019CycleGAN-VC2} and StarGAN-VC2 \cite{kaneko2019stargan}. The generator contains three parts: the downsampling blocks, the bottleneck blocks and the upsampling blocks.
As shown in Figure \ref{fig:stadain3}, the upsampling blocks and the downsampling blocks uses 2D-convolutional layers and instance normalization \cite{ulyanov2016instance} (IN). There are 9 bottleneck blocks,
where each contains a 1D-convolutional layer with the weight adaptive instance normalization (W-AdaIN). The gated linear units (GLU) are used as the activation function.
\subsubsection{Adaptive Instance Normalization}
Adaptive instance normalization \cite{huang2017arbitrary} (AdaIN) was initially proposed for image style transfer tasks. Based on CIN (Equation \ref{sec2:stgan_vc2_cin_layer}), AdaIN uses a speaker encoder to extract the speaker embedding $e_{y} = E(y)$.
\begin{equation}
    AdaIN(f) = \gamma(e_{y}) * (\frac{f - \mu}{\sigma}) + \beta(e_{y}),
\end{equation}
where $AdaIN(f)$ is the output of AdaIN, $e_y$ is speaker embedding, $y$ is target data, $f$ is feature, $\mu$ and $\sigma$ are the mean and the standard deviation of the feature $f$ across time, $\gamma()$ and $\beta()$ are linear functions.
\subsubsection{Weight Adaptive Instance Normalization}
This work tries to improve the data efficiency of the model by using the W-AdaIN module in the bottleneck blocks of the generator.
%
In WAStarGAN-VC, as shown in Figure \ref{fig:w_adain}, the 1D-convolutional weight $w$ has the shape of $[I,J,K]$, where $I$ is the outcoming channel dimensionality of the convolutional layer, $J$ is the incoming channel dimensionality of the convolutional layer, $K$ is the kernel size.

The target speaker data $x_t$ is fed into the speaker encoder to get the speaker embedding $e_{t} = E(x_t)$.
$e_t$ is fed into linear functions to get the affine parameters $\gamma$ and $\beta$.
The affine parameters $\gamma$ and  $\beta$ have the shape of $[B,J]$, where $B$ is the batch size. Then they are expanded on the second and the fourth dimension.
\begin{equation*}
    \gamma_{b,1,j,1}, \beta_{b,1,j,1} = \gamma_{b,j}, \beta_{b,j}
\end{equation*}
Then the weight $w$ is expanded on the first dimension, where $w_{i,j,k}$ is the element of $w$.
\begin{equation*}
    w_{1,i,j,k} = w_{i,j,k} 
\end{equation*}
Next, the expanded weight $ w_{1,i,j,k}$ is transformed by $\gamma_{b,1,j,1}$ and $\beta_{b,1,j,1}$.
\begin{equation}
    w_{b,i,j,k}^* =  \gamma_{b,1,j,1} * w_{1,i,j,k} + \beta_{b,1,j,1}
\end{equation}
The transformed weights $w_{b,i,j,k}^*$ are normalized across the outcoming dimension ($I$).
\begin{equation}
    w_{b,i,j,k}^{**} = \frac{w_{b,i,j,k}^* - \mu_{b,1,j,k}}{\sigma_{b,1,j,k}},
\end{equation}
where $w_{b,i,j,k}^{**}$ is the output of the W-AdaIN module, $\mu_{b,1,j,k}$ and $\sigma_{b,1,j,k}$ are the statistics of $w_{b,i,j,k}^*$ across the outcoming dimension $I$. Finally, the convolution is conducted on feature using the new adapted weight $w^{**}$.
\subsection{Discriminator and Speaker Encoder}
To get speaker-conditioned discriminator output, as in \cite{lee2020many} and StarGAN-V2 \cite{choi2020stargan}, the discriminator uses $N$ parallel speaker-conditioned output layers, where $N$ is the number of the speakers in the training dataset. As shown in Figure \ref{fig:stadain3}, in the discriminator, the first 4 layers are shared across $N$ speakers. For one input sample, the switch selects one of the speaker-conditioned output layers according to the input speaker id. Hence the output of the discriminator is conditioned on the speaker identity.
The speaker encoder also uses the speaker-conditioned parallel output layers. 
Moreover, the speaker encoder uses a statistic pooling layer as in the Xvector \cite{snyder2018x}.
\subsection{Training Objectives}
In WAStarGAN-VC, the training objectives include three parts: adversarial loss, cycle consistency loss and speaker embedding reconstruction loss. As for the adversarial loss, the least square loss \cite{mao2017least} is used, which is the same as in StarGAN-VC2 \cite{kaneko2019stargan}. The cycle consistency loss is the same as in Equation \ref{sec2eq:stgan_vc1_cyc_loss}. 
The speaker embedding reconstruction loss $\mathcal{L}_{spk}$ tries to reconstruct the target speaker embedding $e_t$ from the converted data $y_t$.
\begin{equation}\label{sec3eq:spk_enc_rec_loss}
     \mathcal{L}_{spk} = \mathrm{E}_{x_s, x_t}[||E(x_t) - E(G(x_s, E(x_t)))||_1]
\end{equation}

\section{Experiment Implementation}
\label{sec:4}
The experiments use VCTK \cite{yamagishi2019cstr} dataset \footnote{Source code is available at:  https://github.com/MingjieChen/LowResourceVC, voice samples is available at: https://minidemo.dcs.shef.ac.uk/wastarganvc/}. The VCTK dataset contains English speech studio recordings with 109 speakers. 
The average number of speech samples per speaker is 400.
\subsection{Experiment Setup}
The experiments are split into three situations according to the number of speakers, the number of training samples: (1) for the first situation, 10 speakers with the full training samples are used, (2) for the second situation, 109 speakers with 20 samples per speaker are used, (3) for the third situation, 109 speakers with 5 samples per speaker are used.
StarGAN-VC and StarGAN-VC2 are used as baseline methods. In case that there are no official open source implementations of the StarGAN-VC model and the StarGAN-VC2 model, we implemented two baseline models.
The waveform data is downsampled into 22.05 kHz. Mel-cepstral coefficients (MCEPs) are extracted using PyWorld \cite{morise2016world} toolkit. The StarGAN-based models only focus on the conversion of the MCEPs. 
As in \cite{kameoka2018stargan} and \cite{kaneko2019stargan}, the logarithmic fundamental frequencies (F0s) are transformed linearly.
WORLD \cite{morise2016world} vocoder is used to generate waveform based on the converted MCEPs, the transformed F0s and the aperiodicities (APs). Finally, the loudness of the generated waveform is normalized using PyLoudNorm \cite{pyloudnorm2019} toolkit.
\subsection{Model Configurations}
The proposed WAStarGAN-VC model is implemented using the PyTorch \cite{paszke2019pytorch} toolkit. The optimizer is Adam \cite{kingma2014adam} with the learning rate for the generator and the discriminator as 2e-4 and 1e-4 respectively. 
The MCEPs are randomly cropped into 256-frame segments during training. The batch size is 8 and the training process takes 250k iterations for 30 hours on one single GPU. 
\section{Experiment Results}
The evaluation includes objective evaluation and subjective evaluation. For objective evaluation, we evaluate the models on all three situations. For subjective evaluation, we only evaluate the StarGAN-VC2 model and the WAStarGAN-VC model on the second situation. 
\subsection{Objective Evaluation}
For objective evaluation, as in \cite{arik2018neural}, speaker identification accuracy (ACC) and speaker verification equal error rate (EER) are the measurements of the quality of the converted samples. In this work, a Xvector \cite{snyder2018x} model is pretrained on the VCTK dataset for the whole 109 speakers. The ACC and EER of the converted samples are used as evaluation metrics. For the third situation where the number of the training samples is 5, the StarGAN-VC model failed to generate sensible voices.
\begin{table}[t]\label{tab:spk_cls_results}
    \centering
    \begin{tabular}{c|c|c|c|c|c|c}
    & \multicolumn{2}{c|}{N=10,M=Full} &\multicolumn{2}{|c|}{N=109,M=20,}&\multicolumn{2}{|c}{N=109,M=5,}\\
    & \multicolumn{2}{c|}{S=900}& \multicolumn{2}{|c|}{S=5400}&\multicolumn{2}{|c}{S=5400}\\
    \hline
     Model&ACC&EER &ACC&EER & ACC&EER  \\
     \hline
    StarGAN-VC &64.4&14.88 &54.4&21.96& none & none \\
    StarGAN-VC2 & 91.5&2.99& 79.6&4.61 & 62.6&8.27 \\
    Ours &\textbf{97.0}&\textbf{0.66} & \textbf{95.9}&\textbf{1.77} & \textbf{92.5}&\textbf{3.56} \\
    \hline
    \end{tabular}
    \caption{Objective evaluation results: ACC (\%) denotes speaker identification accuracy, EER (\%) denotes the speaker verification equal error rate. N is the number of speakers, M is the number training samples, S is the number of converted samples for evaluation.}
    \label{tab:obj_results}
\end{table}

As shown in Table \ref{tab:obj_results}, generally, in all three situations, the proposed model yields the best ACC and EER results. For the first situation, WAStarGAN-VC gets ACC 97.0\%, EER 0.66\%. StarGAN-VC2 is slightly worse than WAStarGAN-VC (ACC 91.5\%, EER 2.99\%), StarGAN-VC is much worse (ACC 64.4\%, EER 14.88\%).
For the second situation, WAStarGAN-VC gets ACC 95.9\%, EER 1.77\%. StarGAN-VC2 gets ACC 79.6\%, EER 4.61\%, and StarGAN-VC gets 54.4\%, EER 21.96\%. Both two baseline models are much worse for this situation.
For the third situation, WAStarGAN-VC gets ACC 92.5\%, EER 3.56\%. StarGAN-VC2 gets ACC 62.6\%, EER 8.27\%, which is much worse than the proposed model. 

The objective results show that our proposed model is slightly better than StarGAN-VC2 when using the full of training samples for 10 speakers. However, for the low-resource situations, our proposed model is much better than StarGAN-VC and StarGAN-VC2. This maybe because the proposed model has better data efficiency, which enables it being able to keep the performance under the low-resource situations.
\subsection{Subjective Evaluation}
To assess the naturalness and the similarity, this work conducts the listening tests by comparing WAStarGAN-VC and StarGAN-VC2. The two models are trained under the second situation where the number of speakers is 109 and the number of training samples is 20. 
AB tests are used for the naturalness evaluation, where evaluators need to choose one sample that has better naturalness from two samples generated from two models. For the similarity evaluation, ABX tests are used. Evaluators need to choose one from two samples that is more similar to the real target sample.
A subset of 10 speakers is randomly selected (5 male and 5 female). In total 90 (10*9=90 all conversion directions) samples are evaluated for each model.
\begin{figure}[t]
    \centering
    \includegraphics[width=3.4in]{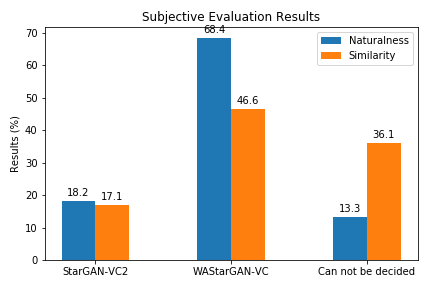}
    \caption{Subjective evaluation results}
    \label{fig:subjective_results}
\end{figure}

As shown in Figure \ref{fig:subjective_results}, for both the naturalness and the similarity, the proposed model obtains the most choices. WAStarGAN-VC gets 68.4\% and 46.6\% choices for the naturalness and the similarity respectively. For the huge gap between WAStarGAN-VC and StarGAN-VC2 on the naturalness, it might because the W-AdaIN module used in the WAStarGAN-VC model alleviates the information loss, therefore the naturalness has gained an improvement.
However, for the similarity, there are 36.1\% of the choices for the option 'can not be decided'.
We compute the correlations of the three choices between the naturalness and the similarity. 
When the naturalness is 'can't be decided', the probability of the similarity being 'can't be decided' is 81.5\%. It means that the naturalness might has correlations with the similarity when the naturalness is low.
\section{Conclusion}
In this work, we proposed the WAStarGAN-VC model and tried to achieve StarGAN-based voice conversion under low-resource situations. 
The subjective and objective evaluation results show that our proposed model has better performance than the baseline models on both naturalness and similarity. Our future work could be one shot voice conversion using StarGAN-based models. 

\vfill\pagebreak



\bibliographystyle{IEEEbib}
\bibliography{strings,refs}

\end{document}